\documentclass[twoside,12pt]{article}\usepackage{macros2e
,din_a4}
\usepackage{epsf}
\newlength{\pcm}
\newlength{\pmm}
\newcommand {\vertex} {\,\epsfxsize=1.5\pcm \parbox{1.5\pcm}{\epsfbox{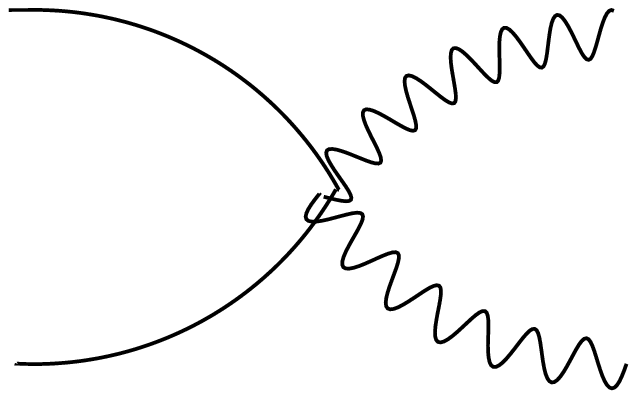}}\,}
\newcommand {\vertexdrei} {\,\epsfxsize=1.5\pcm \parbox{1.5\pcm}{\epsfbox{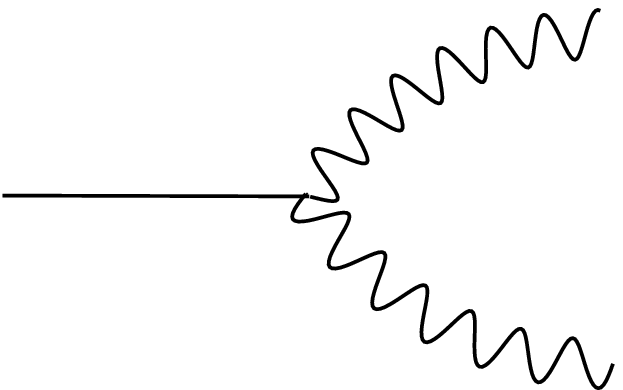}}\,}
\newcommand {\vertexzwei} {\,\epsfxsize=0.75\pcm \parbox{0.75\pcm}{\epsfbox{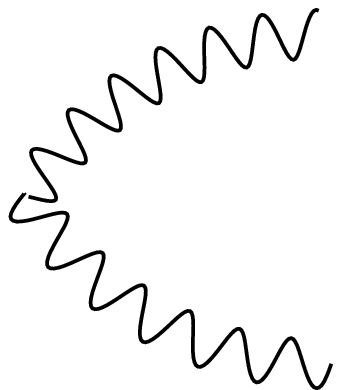}}\,}
\newcommand {\rightvertex} {\,\epsfxsize=0.75\pcm \parbox{0.75\pcm}{\epsfbox{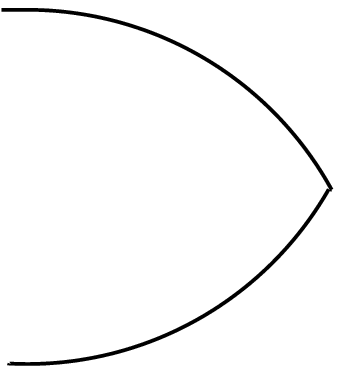}}\,}
\newcommand {\vertexzweip} {\,\epsfxsize=1.5\pcm \parbox{1.5\pcm}{\epsfbox{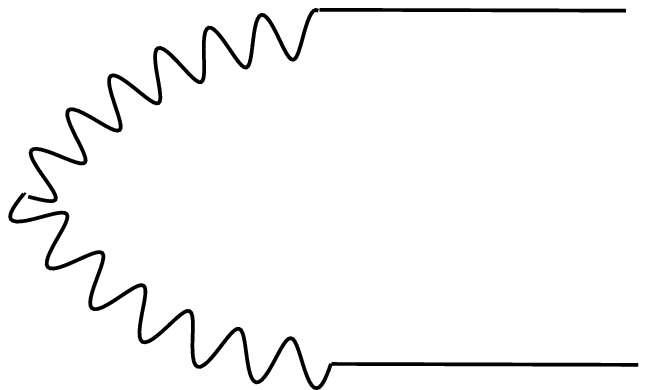}}\,}
\newcommand{\kleb}{\hspace{-2.6\pmm}}
\newcommand {\smallvertex} {\,\epsfxsize=1.0\pcm \parbox{1.0\pcm}{\epsfbox{./KPZ.eps}}\,}
\newcommand {\smallvertexzwei} {\,\epsfxsize=0.5\pcm \parbox{0.5\pcm}{\epsfbox{./KPZ2.eps}}\,}
\newcommand {\smallvertexdrei} {\,\epsfxsize=1.0\pcm \parbox{1.0\pcm}{\epsfbox{./KPZ3.eps}}\,}
\newcommand {\largevertexzwei} {\,\epsfxsize=1.4\pcm \parbox{1.4\pcm}{\epsfbox{./KPZ2.eps}}\,}
\newcommand {\rightlargevertex} {\,\epsfxsize=1.4\pcm \parbox{1.4\pcm}{\epsfbox{./KPZ4.eps}}\,}
\newcommand {\eq}[1]{(\ref{#1})}
\newcommand {\Eq}[1]{Eq.\hspace{0.55ex}(\ref{#1})}
\newcommand {\Eqs}[1]{Eqs.\hspace{0.55ex}(\ref{#1})}
\newcommand{\half}{\frac{1}{2}}
\newcommand{\E}{\varepsilon}
\newcommand{\ts}{
   {\raisebox{-0.5ex}{\parbox{0.5ex}{
      \setlength{\unitlength}{0.5ex}
      \begin{picture}(1,6)
         \thinlines
         \put(0.5,-1){\line(0,1){7}}
      \end{picture}
   }}}
}
\newcommand{\lts}{
   {\raisebox{0ex}{\parbox{0.5ex}{
      \setlength{\unitlength}{0.5ex}
      \begin{picture}(1,11)
         \thinlines
         \put(0.5,-1){\line(0,1){12}}
      \end{picture}
   }}}
}

\newcommand{\rme}{\mbox{e}}
\newcommand{\rmd}{\mbox{d}}
\newcommand{\be}{\begin{equation}}
\newcommand{\ee}{\end{equation}}
\newcommand{\bea}{\begin{eqnarray}}
\newcommand{\eea}{\end{eqnarray}}
\newcommand{\nn}{\nonumber}

\newcommand {\p}{\partial}
\begin{document}
\leftline{\bfseries\Large On the Perturbation Expansion of the {KPZ}-}
\vskip1mm
\leftline{\bfseries\Large Equation}
\vskip6mm
\leftline{\bf\large\normalsize Kay J\"org Wiese%
\footnote{FB Physik, Universit\"at Essen,  45117 Essen,Germany;
e-mail: wiese@theo-phys.uni-essen.de}}
\vskip7mm
\rightline{\begin{minipage}{65ex}\noindent
{\em July 2, 1998} \\
\rule[0.7ex]{65ex}{0.25mm}\\
\small
We present a simple argument to show 
that the $\beta$-function of the $d$-dimensional KPZ-equation 
($d\ge 2$)
 is to all orders in perturbation theory given by
$$
\beta(g_R) = (d-2) g_R - \frac{2}{(8\pi)^{d/2}}\Gamma(2-d/2) g_R^2 \ .
$$
Neither the dynamical
exponent $z$ nor the roughness-exponent $\zeta$ 
have any correction in any order of perturbation theory. 
This shows that standard perturbation theory cannot
attain the strong-coupling regime and in addition breaks down
at $d=4$. We also  calculate a class of correlation-functions
exactly.\\ {\normalsize\rule[0.7ex]{65ex}{0.25mm}}  
\leftline{{\sffamily\bfseries KEY WORDS:} KPZ-equation, growth processes}
\end{minipage}}
\vskip10mm

\section{Introduction} 
During the last years, there has been an increasing interest in 
out of equilibrium dynamics. Among these, a lot of research was 
devoted to
non-linear growth, and 
in particular to the 
Kardar-Parisi-Zhang equation \cite{KPZ}
\bea
\frac{\p h(x, t)}{\p  t} & =& \nu \nabla^2 h(x, t) + \frac\lambda 2 \left( \nabla
h(x, t)\right)^2 + \eta(x, t) \ ,\label{KPZ1} \\
\overline{\eta(x, t) \eta(x', t')} & =& 2D \delta^d(x-x') \delta( t- t') \ .
\label{KPZnoise}
\eea
Thanks to a fluctuation dissipation theorem and the mapping to 
exactly solvable models, 
much is known  for space-dimension $d=1$ \cite{KPZ,HalpinHealyZhang95,Krug97}.
In contrast, 
the   case of $d\ge2$
can  only be attacked by approximative
methods or field-theoretic perturbative
expansions. Using the latter, the fixed point
structure of the renormalization group flow for $d=2+\E$ has been 
obtained \cite{KPZ,FreyTaeuber94,Wiese97c}.
Two domains can be distinguished:
For small effective coupling
\be
        g=\frac{2\lambda^2 D}{\nu^3} \ ,
\ee
the renormalization group flow goes to
0 in the long-wavelength limit. For large coupling the flow
is expected to tend to a
strong coupling fixed point $g=g_{\mbox{\scriptsize sc}}$.
The crossover takes place at $g=g_{\mbox{\scriptsize co}}$,
which turns out to be  of order $\E$ in an $\E$-expansion
and can therefore be studied perturbatively.

In this article we present a  simple argument to resum
the perturbation expansion and to calculate the renormalization 
group functions to all orders. This topic has first been 
adressed in \cite{Laessig95}, but 
is difficult to access there by a non-specialist. The author of the
present publication was therefore encouraged to find a simple derivation,
 which sets the results  on a clear footing and allows to study the
limits of the method. Emphasis is laid upon 
a pedagogical presentation, understandable 
with an elementary background in renormalization theory. 
We will therefore perform all steps of the renormalization program by
using elementary tools only. This includes a proof of 
perturbative  renormalizability, 
which in standard field theories is a formidable task, see e.g.~\cite{Zinn}
and references therein. 
 
Let us also mention that similar conclusions have independently 
been obtained by H.K.~Janssen \cite{JanssenPrivate}, and in  a different context and with completely different methods in \cite{ImbrieSpencer1988,%
CookDerrida89}.

\markboth{K. Wiese}{On the Perturbation Expansion of the {KPZ}-Equation}
\section{Summation of the KPZ-equation to all orders in 
perturbation theory}
\label{Solution of the KPZ-equation to all orders in 
perturbation theory}
First of all, we want to eliminate the nonlinear term in \Eq{KPZ1}. Using the well-known Cole Hopf transformation 
\begin{equation} \label{Cole Hopf}
W(x,{t}):= \rme^{\frac\lambda{2\nu} h(x,{t})} \ ,
\end{equation}
and absorbing a factor of $\nu$ into $t$ leads 
to the following equations in terms of $W(x,t)$:
\begin{eqnarray}
\frac{\p}{\p t} W(x,t)& =&  \Delta W(x,t)  + \frac{\lambda} {2 \nu^2} \eta(x,t) W(x,t) \ ,
 \label{KPZ7} \\
\overline{\eta(x,t) \eta(x',t')} & =& 2\nu D \delta^d(x-x') \delta( t- t')\ .
\end{eqnarray}
In interpreting \Eqs{KPZ1} and \eq{KPZ7} in It\^o-discretization, 
we have explicitly subtracted a drift term $\sim W(x,t)$. Thus the 
expectaton value of $W(x,t)$ will be constant.

We are now in a position to write down the generating functional 
for the dynamic expectation values
\begin{equation} \label{gen functional}
\int {\cal D}\left[\tilde{W}\right] {\cal D}\left[W\right]
{\cal D}\left[\eta\right]
 \rme^{-{J} \left[W, \tilde{W}, \eta\right] + \int j (x, t) W (x, t) + \tilde{\mbox{\small \it \j}}(x, t) \tilde W (x, t)}
\end{equation}
with
\bea 
{J}\left[W,\tilde{W},\eta\right] = 
\int _{x, t}\bigg[ \tilde{W} (x, t) \bigg( \dot{W}(x, t) \!\!&-&\!\!  \Delta W (x, t) 
- \frac {\lambda} {2\nu^2} \eta (x, t) W (x, t) \bigg)
\nn\\
&+&\!\! \frac {1} {4 D \nu} \eta^2(x, t) \label{gen func 1} \bigg]\ .
\eea
Expectation values are obtained from \Eq{gen functional}
through variation with respect to $j (x, t)$ and $\tilde{\mbox{\it \j}}(x, t)$. 
The interpretation of the functional \eq{gen func 1} 
is simple: The term proportional 
to $\tilde W(x,t)$ is just the equation of motion \eq{KPZ7}, thus  
integrating over all purely imaginary fields forces the 
equation of motion to be satisfied. The last term in 
\Eq{gen func 1} is the noise distribution.
Note that the path integral runs  over positive values of 
$W (x, t)$ only, since $W (x, t) = \rme^{\frac\lambda{2 \nu} h(x, t)}$. 

The noise-integration can be done. We obtain a simplified action 
\bea 
J\left[W, \tilde W\right] &=& \int_{x, t} \tilde{W} (x, t) \left( \dot{W} (x, t)
- \Delta W (x, t)\right) 
- \frac{g}2 \left( \tilde{W} (x, t) W (x, t)\right)^2 \ ,\label{eff-func}
\nn\\&&
\eea
where 
\be
g = \frac {2\lambda^2 D} {\nu^3} \ .
\ee
As a side remark, let us note that
another way to obtain \Eq{eff-func} is to write down the generating
functional  for the original KPZ-equation \eq{KPZ1} and then to perform 
a change of coordinates \cite{JanssenPrivate}
\bea \label{canonical transform}
W(x,t)&:=& \rme^{\frac\lambda{2\nu} h(x,t)}\ , \nn\\
\tilde W(x,t)&:=& \tilde h(x,t)\, \rme^{-\frac\lambda{2\nu} h(x,t)} \ .
\eea
This transformation  leaves 
the integration measure invariant. 

\Eq{KPZ7} only makes sense when specifying the initial conditions,
i.e.\ $W(x,t)$ at time $t=0$. The simplest choice $W(x,0)=0$ leads to 
$\partial_t W(x,0)=0$ and consequently to $W(x,t)\equiv 0$.
We therefore start with 
\be \label{Winitial}
W(x,0)=1 \ ,
\ee
which is equivalent to a flat initial condition for $h(x,t)$, 
namely $h(x,0)=0$. In order to eliminate the constant
part of \Eq{Winitial} from perturbation theory, we set
\be
 W(x,t) = 1+ w(x,t) \ .
\ee 
The response-function of the non-interacting theory (``free response-function'') is
\begin{eqnarray}
R(x-x', t-t') &=&  \langle w(x, t) \tilde W(x',t') \rangle_{0} \nonumber \\
& =&  \Theta(t-t') \left[4\pi (t-t')\right]^{-d/2} \rme^{-(x-x')^2/4(t-t')} \ .
\hspace{1cm}
\label{free-resp}
\end{eqnarray}
All other free expectation values vanish
\bea \label{WW}
 \langle w(x, t) w (0,0)\rangle_{0} &=& 0 \ ,\nn\\
\label{tilde W tilde W}
\langle \tilde{W} (x, t) \tilde {W} (0,0) \rangle_0 &=& 0 \ .
\eea

Let us now adress the problem of restricting the path-integral to values of $W(x, t) > 0$. Starting with $W (x, t) > 0$, the time evolution in \Eq{KPZ7} will keep $W (x, t) > 0$ for all $t$. This is easily verified for
 vanishing noise, therefore the free response-function 
\eq{free-resp} is correct. We shall see below that also perturbation theory respects this property. 

Perturbation theory is developed by starting from the functional \eq{eff-func}. The non-linear term is
\begin{equation} \label{vetex}
\half \left( \tilde{W} (x, t) W (x, t)\right)^2 
=\half \tilde W^2(x, t) +\tilde W^2(x, t)  w(x,t) +
\half \left( \tilde{W} (x, t) w (x, t)\right)^2 \ ,
\end{equation}
and we denote 
\bea
\half \int_{x,t} \tilde W^2(x, t) &=& \vertexzwei \ ,\nn\\
\half \int_{x,t} \tilde W^2(x, t)  w(x,t) &=& \vertexdrei\ , \label{vertices}\\
\half \int_{x,t} \left( \tilde{W} (x, t) w (x, t)\right)^2 &=& \vertex \ .\nn
\eea
Since $\smallvertexzwei$ and $\smallvertexdrei$ by its own can not
build up divergent diagrams, we neglect them for the moment
and start by analysing 
 the perturbative expansion of an observable $O$ with $\smallvertex$
only 
\begin{equation}
\langle O \rangle = \langle O\, \rme^{ g \smallvertex}\rangle_0
\ .
\end{equation}
The basic ingredient is the exponential of the interaction
\begin{equation} \label{expofint}
\rme^{g \smallvertex} \ ,
\end{equation}
from which we have to build  vertices in perturbation theory. 
First of all, there is no vacuum-correction, as self-contractions of $\smallvertex$ vanish identically due to causality. This also holds for the contraction of more than one vertex. With the same argument, we conclude, that no diagram with two external legs can be constructed. Therefore, there is no divergent contribution to both $\tilde{W}€\dot{W}\equiv 
\tilde{W}€\dot{w} $ and $\tilde{W} \Delta W\equiv 
\tilde{W}€\Delta{w} $ at any order in perturbation-theory, therefore  $\nu$ (hidden in $t$) has not to be renormalized\footnote{Note that this is 
not in contradiction with the non-trivial value for $z$ obtained 
in dimension
$d=1$: there, as well as for $d\ge2$, the 
non-trivial fixed point describing the rough phase
 is in the strong coupling regime, i.e.\ {\em
not} acceccible by a systematic
perturbation expansion; this means  that the 
 expansion parameter is always large and the expansion uncontrolled.}. The only possible diagrams are chains of $\smallvertex$, of the form $\smallvertex\kleb\smallvertex$, $\smallvertex\kleb\smallvertex\kleb\smallvertex$ and so on or higher order
vertices. The latter are irrelevant in perturbation theory \cite{Zinn}.

We  therefore write
\bea  \label{ladder}
\rme^{g \smallvertex} = 1 &+& g \vertex + g^2 \vertex\kleb\vertex \nn\\
&+& g^3 \vertex\kleb\vertex\kleb\vertex  
\label{pert-exp}
  + g^4 \vertex\kleb\vertex\kleb\vertex\kleb\vertex + \ldots \nn\\
 &+& \mbox{higher order vertices} \ ,
\eea
where the time-argument of the vertices grows from left to right. Note that the combinatorial factor of $\frac {1} {n!}$ which comes from the expansion of the exponential function at order $g^n$  has canceled against the $n!$ possibilities to order the vertices in time. 
In addition, any bubble appears with a combinatorial factor of 2, which 
cancels against factors of $1/2$ from the vertex, \Eq{vetex}. So 
any of the chain diagrams in \eq{expofint} still contains a factor of 
$1/2$. 

To proceed further, we first suppress the ``higher
order vertices'' in \Eq{pert-exp}, as the only divergencies they may
contain are sub-chains as those depicted in \Eq{pert-exp}, that will be 
treated here.

Second, we can switch to Fourier-representation, thus
regard the diagrams in \Eq{ladder} as a function 
of the external momentum $p$ and
frequency $\omega$ instead of the coordinates $x$ and $t$, 
and finally integrate over $p$ and $\omega$ instead of $x$ and $t$.
Then, each chain in \Eq{ladder} factorizes, i.e.\ can 
be written as product of the vertex $\smallvertex$ 
times a power
of the elementary loop diagram (which is a function of $p$ and
$\omega$) 
\begin{equation} \label{fac-prop}
\stackrel{p,\omega}\longrightarrow
\underbrace{\vertex\kleb\vertex\kleb\vertex\kleb\vertex\kleb\vertex}_%
{\mbox{\scriptsize $n$ loops}}
 = \left( \vertexzwei\kleb\rightvertex_{p,\omega} \right)^n   \vertex \ .
\end{equation}
\Eq{ladder} is a geometric sum, equivalent to
\be
1 + g \,\frac1{1-g\left. \vertexzwei\kleb\rightvertex_{p,\omega} \right.} 
\,\vertex \ ,
\ee
and one reads off the effective 4-point function 
\be \label{eff Gamma 4}
\Gamma_{ww\tilde W \tilde W}\ts_{p,\omega} =  
g \,\frac1{1-g\left. \vertexzwei\kleb\rightvertex_{p,\omega} \right.} \ .
\ee
As we shall show below, the loop integral in 
\Eq{eff Gamma 4} is divergent for any 
$p$ and $\omega$ when $d€\to 2 $. 
Renormalisation means to absorb this divergence into
a reparametrization of the coupling constant $g$: 
We claim
 that there is a function $a=a(d)$, such that the 4-point function
 is finite (renormalized) 
as a function of $g_R$ instead of $g$, when setting
\begin{equation} \label{ren-bare}
g = Z_g g_R\mu^{-\varepsilon}
\end{equation}
with 
\begin{equation} \label{Zg}
Z_g = \frac {1} {1+a g_R} \ , \quad \E=d-2 \ .
\end{equation}
$\mu$ is an arbitrary scale, the so-called renormalization scale. 
As a function of $g_R$, the 4-point function reads 
\be \label{Gamma4R}
\Gamma_{ww\tilde W \tilde W}\ts_{p,\omega} =
	\frac{g_R\mu^{-\E}}{1+\left(  a  -
\mu^{-\E}\,\vertexzwei\kleb\rightvertex_{p,\omega} 
 \right) g_R} \ .
\ee
To complete the proof, we have to calculate  the elementary diagram, 
\be
\stackrel{p,\omega}\longrightarrow \ 
\raisebox{1.1\pcm}[1.5\pcm][0mm]{\parbox{0mm}{$\hspace{-.1\pcm}{\frac \omega 2 +\nu,\ \frac p2+k\atop}$}}
\raisebox{-1.4\pcm}[-1.6\pcm][0mm]{\parbox{0mm}{$\hspace{-.1\pcm}{\frac \omega 2-\nu,\ \frac p2-k\atop}$}}
\largevertexzwei\kleb\hspace{-0.18mm}\rightlargevertex
\ 
\stackrel{p,\omega}\longrightarrow  \ \ \ \ \ .
\ee
This is 
\be
\int\frac{\rmd^d k}{(2\pi)^d} \int\frac{\rmd \nu}{2\pi} \frac{1}{\left(\frac p2+k\right)^2 +i \left( \frac \omega 2
+\nu\right)}
\frac{1}{\left(\frac p2-k\right)^2 +i \left( \frac \omega 2
-\nu\right)}  \ .
\ee
To perform the integration over $\nu$, the integration path can be 
closed either in the upper or lower half-plane. Closing it in 
the upper half-plane, we obtain:
\bea
\int\frac{\rmd^d k}{(2\pi)^d}  
\frac{1}{\left(k+\frac p2\right)^2 +\left(k-\frac p2\right)^2+i \omega}
&=&\int_0^\infty \rmd s\int\frac{\rmd^d k}{(2\pi)^d}  \rme^{-s\left(2 k^2 +\half p^2 +i\omega \right) } 
\nn\\
& &\hspace{-3cm}=\frac{1}{(8\pi)^{d/2}} \int_0^\infty \rmd s\, s^{-d/2}\rme^{-s( \half p^2 +i\omega )}
\nn\\
& &\hspace{-3cm}=\frac{1}{(8\pi)^{d/2}} \left(\half p^2 +i\omega\right)^{d/2-1}
\Gamma\left(1-\frac d2\right) \ . \label{div}
\eea
The 4-point function in \Eq{Gamma4R} therefore
depends on $p$ and $\omega$, and more specifically on the combination 
$\half p^2 +i\omega$. 
We now chose a subtraction scale, i.e.\ we demand 
that $\Gamma_{ww\tilde W \tilde W}$ evaluated at 
$\mu^2=\half p^2 +i\omega$ be
\be \label{ren cond}
\Gamma_{ww\tilde W \tilde W}\ts_{  \half p^2 +i\omega=\mu^2}=g_R \ .
\ee
This is achieved by setting 
\be \label{a}
a= \frac{1}{(8\pi)^{d/2}} \Gamma\left(1-\frac d2\right) 
\equiv \frac{2}{(8\pi)^{d/2}} \Gamma\left(2-\frac d2\right) \frac1\E \ .
\ee
Moreover, since 
\be
\frac1\E  \left(\half p^2 +i\omega\right)^{d/2-1} \mu^{-\E} 
\ee
is finite in the limit $\E\to0$ as long as the  
combination of $\half p^2 +i\omega$ is finite, 
it can be read off from \Eq{Gamma4R} that then also 
$\Gamma_{ww\tilde W \tilde W}\ts_{p,\omega}$ is finite. 
(If useful, either $p=0$ or $\omega=0$ may safely be  taken.)
This completes the proof. Note that this ensures that the 
model is renormalizable to all orders in perturbation-theory,
what is normally a formidable task to show \cite{Zinn}.

The $\beta$-function that  we shall calculate now is
exact to all orders in perturbation theory. As usual, it is defined
as the variation of the renormalized coupling constant, keeping the 
bare one fixed
\be
	\beta(g_R) = \mu \frac{\partial}{\partial \mu}\lts_{g} g_R \ .
\ee
From \Eq{ren cond} we see that it gives the 
dependence of the 4-point function on $p$ and $\omega$ for fixed bare
coupling.
Solving 
\be
g= \frac{g_R \mu^{-\E}}{1+ a g_R}
\ee
for $g_R$, we obtain
\be
g_R=\frac{g}{\mu^{-\E}- a g} \ ,
\ee 
and hence
\be
\beta(g_R) =\E g_R (1+ a g_R) \ .
\ee
Using $a$ from \Eq{a}, our final result is
\be \label{final result}
\beta(g_R)= (d-2) g_R - \frac{2}{(8\pi)^{d/2}} \Gamma\left(2-\frac d2\right) g_R^2 \ ,
\ee
as stated in the abstract. It shows that 
standard
perturbation theory fails to produce a strong coupling fixed
point, a result which cannot be overemphasized. 
This means that any treatment of the strong coupling regime
has to rely on non-perturbative methods. It does of course
not rule out the possibility  to find an exactly solvable 
model, non-equivalent to KPZ, for which it is possible 
to expand towards the strong-coupling regime of KPZ. 
Note that also for $d=1$ this equation does not possess a 
fixed point describing the rough phase; the latter is in the
strong coupling regime, {\em not} accessible by a perturbation 
expansion.

Let us also note that the
 $\beta$-function is divergent at $d=4$, and therefore 
our perturbation expansion breaks down at $d=4$.
To cure the problem, a lattice regularized version of 
\Eq{KPZ7} may be used. However, then the lattice 
cut-off $a$ will enter into the equations and the result 
is no longer model-independent. 
This may be interpreted as  $d=4$ being the upper
critical dimension of KPZ, or as sign for
a simple technical problem.  Compare also \cite{BundschuhLaessig96}.

\section{Critical exponents and connection to calculations in the standard 
representation of the KPZ-equation}\label{Connection ...}
Quite a lot of work has been done by directly working in
the KPZ-picture.  
A crucial point to understand is therefore the 
relation of the KPZ-picture and the Cole-Hopf transformed
model used here. The relation is easy on the level of 
the $\beta$-functions, which are the same for both models. 
In the Cole-Hopf picture  it is also easy  to see that,
as there are no corrections to the response function, $\nu$
and by this means
the dynamical exponent $z$ does not acquire any perturbative
correction, and thus 
\be \label{z=2}
	z=2
\ee
to all orders.

What is not so easy to compare are 
correlation functions of the height-variable $h(x,t)$
in the original KPZ-language with objects of the 
Cole-Hopf transformed theory. One has to 
compare, see \Eq{Cole Hopf}
\be
\left<  h(x,t)\, h(x',t') \right> \sim 
\left< \ln W(x,t)\, \ln W(x',t') \right> \ .
\ee
As logarithms are difficult to handle, one can also 
study expectation values of vertex-operators 
\be \label{vertexoperator}
\left<
\rme^{\kappa W(x,t)} \rme^{\rho W(x',t')}  \right>
\label{genW}
\ee
with arbitrary $\kappa$ and $\rho$. 
Expectation values of $h$ can be reconstructed by using 
the identity
\be \label{integral-transform}
\ln W = \lim_{\delta \to 0} \frac1\delta
\left[1- W^{-\delta}\right]  \nn\\
= \lim_{\delta \to 0}\int\limits_0^{\infty}\frac {\rmd \kappa}\kappa
\kappa^{\delta}\left[\rme^{-\kappa}-\rme^{-\kappa W}\right]\ .
\ee
Therefore the characteristic functions 
contain much more information than e.g.\ the 2-point
function
\be
\left<  W(x,t)\,W(x',t') \right>  \ .
\ee

Using the tilt-invariance of the KPZ-model 
\cite{MedinaEtAl89}, \Eq{z=2} also implies that the roughness exponent $\zeta_h$,
defined via  (for $\zeta_h<0$; for  $\zeta_h>0$ one would use
$\left< \left( h(x,t)-h(x',t) \right)^2 \right>$ instead)
\be
\left<  h(x,t)\,h(x',t) \right> \sim |x-x'|^{2 \zeta_h}
\ee
of the $h$-field 
at a non trivial fixed point is
\be
\zeta_h=0\ .
\ee 
To study correlation-functions more
 directly, we use the first term from \Eq{vertices} to calculate 
\bea
 \left< w(x,t)\, w(x',t') \right> &=& g \int_{y,\tau} \ {\scriptstyle y,\tau}\vertexzweip
{{\scriptstyle x,t} \atop {\scriptstyle x',t'} } \nn\\
&=&g\int_{y,\tau} R(x-y,t-\tau) R(x'-y, t'-\tau) \ .
\label{new}
\eea
Note that this is the only diagram which contributes, since more
complicated diagrams involving loops like $\vertexzwei\kleb\vertex$,
have to be taken at zero momentum and frequency
 and thus vanish according to \Eq{div}.

Further, when $t$ and $t'$ are small, the expectation value in 
\Eq{new} is small, too. This is physical, since the surface has not 
much grown yet. In the other limit of large times, $t,t' \to \infty$ and
keeping $t-t'$ fixed, the r.h.s.\ of \Eq{new} converges towards 
\be
g\, C(x-x',t-t') \ ,
\ee
where $C(x,t)$ is the standard dynamic correlation function
which in Fourier space reads
\be
C(k,\omega)= \frac1{k^4+\omega^2} \ .
\ee
For equal times, this relation reads
\be
\left< w(x,t)\, w(x',t) \right> \sim g\, |x-x'|^{2-d}=g_R Z_g  |x-x'|^{2-d}
\ ,
\ee
leading to a renormalization for $w$ of the form 
\be
w_R=Z_g^{-\half}w
\ee
and to a roughness exponent $\zeta_w$ for the field $w$ in 
\be
\left< w(x,t)\, w(x',t) \right> \sim |x-x'|^{2\zeta_w}
\ee
with 
\bea
\zeta_w&=&\zeta_0 +\delta \zeta_w \ ,\nn\\
\zeta_0&=& \frac{2-d}2 \ , \\
\delta \zeta_w&=&\half \mu \frac{\partial}{\partial \mu}\ln Z_g \ .\nn
\eea
Using \Eqs{ren-bare} and \eq{Zg}, we can express $\beta(g)$ in terms 
of $\delta \zeta_w(g_R)$ as
\be
\beta(g_R) = g_R\left[ \E -2\delta \zeta_w \right] \ .
\ee
At a non-trivial fixed point $g_R=g^*>0$, i.e.\ at the roughning
transistion, this induces 
\be
\delta \zeta_w(g_R) =\frac\E2 \ ,
\ee
and therefore
\be
\zeta_w = 0 \ .
\ee
It is tempting to indentify $\zeta_w$ with $\zeta_h$. A priori, 
this is surprising, since $h$ and $w$ are very different
observables, related by 
\be \label{related}
w(x,t)=\rme^{\frac\lambda\nu h(x,t)}-1 \ .
\ee
The putative identity $\zeta_w=\zeta_h$ will therefore
only hold, if $w$ is, within correlation functions, well approximated 
by the term linear in $h$ on the r.h.s.\ of \Eq{related}. It will 
certainly break down when $\zeta_h>0$, i.e.\ in the 
strong-coupling regime. It is still possible to relate correlation-%
functions for $h$ and $w$ via integral-transforms as in 
\Eq{integral-transform}, as long as  the expectation value in \Eq{vertexoperator} is dominated by contractions with $\smallvertexzwei$
only, leading to purely Gaussian correlations. 
However, in the strong-coupling regime,  
also non-linear 
terms, of which the first is $\left< w^2(x,t) w(x',t') \right>$,
will contribute to \Eq{genW}, making a systematic analysis easier 
said than done.
This is another difficulty of  the 
perturbation expansion beyond the 
roughning transition. 

On a more technical level, it is worth realizing that the  above relations 
can be used to  simplify the perturbation expansion in the original 
KPZ-language.

\section{Conclusions}
In this article, we have presented  a simple method to 
resum the perturbation expansion of the KPZ-equation and
to calculate the renormalization group $\beta$-function
 to all orders in perturbation theory, including a proof
of perturbative renormalizability. The main conclusion
is that there is no anomalous contribution to the dynamical
exponent $z$ in the weak-coupling regime and at the roughening
transition. We also have given some indications of 
why standard perturbation theory fails 
in describing  the strong-coupling fixed regime.

\section*{Acknowledgements}
It is a pleasure to thank H.W.\ Diehl, H.K.\ Janssen, H.\ Kinzelbach
and  J.\ Krug for useful discussions.

\bibliography{../citation/citation}
\bibliographystyle{KAY}

\end{document}